# Walking vector soliton caging and releasing

Yaroslav V. Kartashov, Victor A. Vysloukh, and Lluis Torner

*ICFO-Institut de Ciencies Fotoniques, and Universitat Politecnica de Catalunya, Mediterranean Technology Park, 08860 Castelldefels (Barcelona), Spain*

We address the formation and propagation of vector solitons in optical lattices in the presence of anisotropy-induced walk-off between ordinary and extraordinary polarized field components. Stable vector solitons trapped by the lattice form above a threshold power, while decreasing the lattice depth below a critical value results in the abrupt release of the caged solitons, that then move across the lattice and may get trapped in a desired lattice channel.

OCIS codes: 190.0190, 190.6135

Optical solitons may appear as scalar (i.e., single-field) or as vector entities, where two field components mutually trap together to form a single localized state [1]. Vector solitons are encountered in a variety of physical settings [2-12]. The field components forming vector solitons may experience temporal [2,4,7-9] or spatial [12] walk-off. In this case walking vector solitons form when the pulses or beams lock together in spite of the linear drift [7]. Vector solitons may form not only in uniform materials, but also in waveguide arrays or lattices [13-17]. However, to date vector lattice solitons have been addressed only in settings without spatial walk-off between the field components.

   In this Letter we consider vector solitons in optical lattices imprinted in anisotropic Kerr media in the presence of Poynting-vector walk-off. Our motivation is to elucidate the interplay between walk-off that causes soliton motion and the refractive index modulation that restricts soliton mobility [18-23]. We show that, above a threshold power, lattices support stable elliptically polarized vector solitons that are trapped in a given lattice channel in spite of the underlying walk-off. Decreasing the lattice depth results in the release of the solitons, that then walk across the lattice and may get trapped in a desired lattice channel.

We describe the propagation of two coherently interacting ordinary ($x$) and extraordinary ($y$) polarized waves in a birefringent medium with an imprinted optical lattice with the nonlinear Schrödinger equations for the dimensionless field amplitudes $q_x$ and $q_y$:

$$i\frac{\partial q_x}{\partial \xi} = -\frac{1}{2}\frac{\partial^2 q_x}{\partial \eta^2} - i\alpha\frac{\partial q_x}{\partial \eta} - q_x\left(|q_x|^2 + \frac{2}{3}|q_y|^2\right) - \frac{1}{3}q_x^*q_y^2\exp(-i\beta\xi) - pR(\eta)q_x,$$
$$i\frac{\partial q_y}{\partial \xi} = -\frac{1}{2}\frac{\partial^2 q_y}{\partial \eta^2} - q_y\left(|q_y|^2 + \frac{2}{3}|q_x|^2\right) - \frac{1}{3}q_y^*q_x^2\exp(+i\beta\xi) - pR(\eta)q_y. \quad (1)$$

Here $\eta, \xi$ are the transverse coordinate and propagation distance normalized to the beam width and diffraction length, respectively; $p$ is the refractive index modulation depth and the function $R(\eta) = \cos(\Omega\eta)$ describes transverse shape of the lattice. We consider a general case of off-axis propagation relative to the crystal optical axis, so that the spatial walk-off $\alpha$ determined by the angle between Poynting vectors of $x$ and $y$ components has to be taken into account. Without loss of generality we set the phase mismatch $\beta = 3$ ($\beta$ affects the energy exchange between components) and the lattice frequency $\Omega = 4$.

First, we address the properties of stationary vector solitons supported by the optical lattice in the presence of walk-off. Such stationary solutions exist only for zero transverse velocity (i.e. they are caged in a lattice channel) and have the form $q_x = (u_x + iv_x)\exp(ib\xi)$ and $q_y = (u_y + iv_y)\exp[i(b + \beta/2)\xi]$, where $b$ is the propagation constant, while $u_{x,y}(\eta)$ and $v_{x,y}(\eta)$ are real and imaginary parts of the corresponding fields. Once stationary solutions of Eqs. (1) are obtained, we analyze their stability by adding small perturbations, linearizing Eqs. (1) around the stationary solutions, and solving the resulting linear eigenvalue problem for the perturbation profiles and complex growth rates $\delta = \delta_r + i\delta_i$.

When $\beta > 0$ Eqs. (1) have three types of solutions: $q_x \neq 0$, $q_y = 0$ ("slow" scalar mode polarized along $x$ axis); $q_x = 0$, $q_y \neq 0$ ("fast" scalar mode polarized along $y$ axis); and $q_x \neq 0$, $q_y \neq 0$ (elliptically polarized vector mode). A typical example of elliptically polarized vector solitons is shown in Figs. 1(a) and 1(b). Due to the presence of walk-off and four-wave-mixing, both $x$ and $y$ components exhibit a spatially chirped phase-front, a feature characteristic of walking solitons [24]. The intensity distributions are modulated due to the presence of the lattice. This modulation becomes more pronounced for broad low-power solitons, while high-amplitude solitons concentrate in a single lattice site. The energy flow $U = U_\text{x} + U_\text{y} = \int_{-\infty}^{\infty} [(u_\text{x}^2 + v_\text{x}^2) + (u_\text{y}^2 + v_\text{y}^2)] d\eta$ of elliptically polarized solitons is a monotonically increasing function of $b$ [Fig. 1(c)]. Vector solitons bifurcate from "fast" $y$-polarized modes. The fraction of energy $S_\text{x,y} = U_\text{x,y}/U$ carried by $x$ and $y$ components as a function of $b$ is shown in Fig. 1(d). Elliptically polarized solitons exist for $b \geq b_\text{co}$ and for energy flows above a threshold. The cutoff $b_\text{co}$ grows monotonically with increasing walk-off [Fig. 1(e)] and lattice depth. The threshold energy flow also grows with $\alpha$. The linear stability analysis indicates that "slow" $x$-polarized modes are always stable, while "fast" $y$-polarized mode becomes unstable for $b \geq b_\text{co}$ after bifurcation point [Fig. 1(f)]. Elliptically polarized vector solitons exhibit complex $\delta_r(b)$ dependence with several stability domains.

Even though the vector solitons are a locked state of two field components that experience walk-off, rigorous stationary walking soliton solutions do not exist in the presence of the lattice. This is a consequence of the broken transverse symmetry of the periodic refractive index modulation. However, under appropriate conditions, vector solitons do walk across the lattice (Fig. 2). The central motivation of this Letter is to elucidate the conditions at which the mutual dragging induced by the walk-off becomes dominant over the caging effect of the lattice, so that vector solitons are released. The phenomenon is best illustrated by taking as input vector solitons supported by a given lattice, and then study soliton propagation when decreasing the lattice depth. Such decrease causes a strong energy exchange between $x$- and $y$-components resulting in enhancement of energy fraction carried by the component affected by the walk-off. Hence, the dragging force pulling solitons away from the input channel is enhanced so that below critical lattice depth solitons escape and start walking across the lattice. Because the walking solitons leak energy when they cross lattice channels, they can be eventually trapped in a different lattice channel (Fig. 2). Thus, input solitons can be routed to desired output channels by varying the lattice depth. Although soliton releasing and trapping is possible with scalar fields propagating across the lattice because of an initial phase tilt, the vectorial interactions are accompanied by energy exchange between the field components thus enriching the opportunities to control the output soliton position by varying, e.g., the input power carried by each field component. Note also the fundamental difference existing between an input beam with an initial linear tilt in a single field and the nonlinear mutual dragging caused by walk-off in vector solitons.

Figure 3 shows the output lattice channel where walking soliton is located at $\xi = 50$ as a function of lattice depth for several values of the input power. The inputs in all cases correspond to the elliptically polarized vector solitons supported by a lattice with $p = 4$. The vector solitons start walking across the lattice at $p < p_{\text{cr}}$. The critical lattice depth rapidly decreases with increasing $U$ [Fig. 4(a)], since corresponding Peierls-Nabarro potential barrier grows for high-amplitude solitons [22]. Soliton release cannot take place for too small energy flows, because for the input vector states the amplitude of $x$-component decreases rapidly with decreasing $U$ [Fig. 1(d)]. When such solitons with $U_{\text{x}} \ll U_{\text{y}}$ are used as input, they remain immobile even in shallow lattices. Under proper conditions one can find intervals of lattice depths corresponding to routing into channels with progressively increasing numbers [Figs. 3(a) and 3(b)]. At high powers the dependence $n_{\text{out}}(p)$ may become irregular [Figs. 3(c) and 3(d)].

Since in actual experiments the shape of input beam is usually far from the shape of exact vector soliton it is important to elucidate whether soliton release can be achieved with input beams having arbitrary shapes, such as e.g., $q_{\text{x,y}}|_{\xi=0} = A \operatorname{sech}(A\eta)$. Figure 2(b) that shows propagation trajectories for such input for different lattice depths at $A = 1.5$ confirms that this is the case, while Fig. 4(b) shows corresponding critical lattice depth versus input energy flow. Smaller radiation for sech input is due to the fact that for $\beta > 0$ in exact vector solitons the power carried by the $x$ component (that is subjected to walk-off) is always smaller than the power carried by the $y$ component. This results in additional energy transfer from $y$ to $x$ component accompanied by stronger radiation than in the case of sech input where $S_{\text{x}} = S_{\text{y}}$.

Summarizing, optical lattices imprinted in anisotropic Kerr media support stable elliptically polarized vector solitons caged at a lattice channel in the presence of walk-off. Vector solitons can walk across the lattice when the dragging induced by the walk-off overcomes the trapping induced by the lattice. Vector walking soliton caging and releasing is possible not only in harmonic lattices, but in other periodic refractive index landscapes, e.g. in arrays of evanescently coupled Gaussian waveguides.

# References with titles

# References without titles

**Figure captions**

Figure 1. Profiles of (a) $x$ and (b) $y$ components of elliptically polarized vector soliton at $b = 4.2$, $\alpha = 1.4$. In gray regions $R(\eta) \geq 0$; in white regions $R(\eta) < 0$. (c) $U$ versus $b$ for $x$-, $y$-, and elliptically polarized solitons at $\alpha = 1.4$. (d) $S_x$ and $S_y$ versus $b$ at $\alpha = 1.4$. Points in (c), (d) correspond to solitons in (a), (b). (e) Cutoff for existence of elliptically polarized soliton versus $\alpha$. (f) $\delta_r$ versus $b$ for $y$- and elliptically polarized solitons at $\alpha = 1.4$. In all cases $p = 4$.

Figure 2. (a) Dynamics of $x$ component at $p = 0.50$ (1), 0.36 (2), and 0.22 (3) when input beam corresponds to vector soliton obtained at $b = 6$, $p = 4$. (b) The same as in (a) but for sech-shaped input beams and $p = 0.60$ (1), 0.32 (2), and 0.23 (3). Distributions corresponding to different $p$ are superimposed. In all cases $\alpha = 1.4$.

Figure 3. The number of the output channel versus $p$ at $\alpha = 1.4$ for elliptically polarized input vector soliton with (a) $U = 6.38$, (b) 6.78, (c) 6.98, and (d) 7.15 obtained at $p = 4$. Vertical dashed lines indicate critical lattice depth.

Figure 4. $p_{\text{cr}}$ versus $U$ for (a) elliptically polarized input vector soliton and (b) sech-shaped input beams at $\alpha = 1.4$.

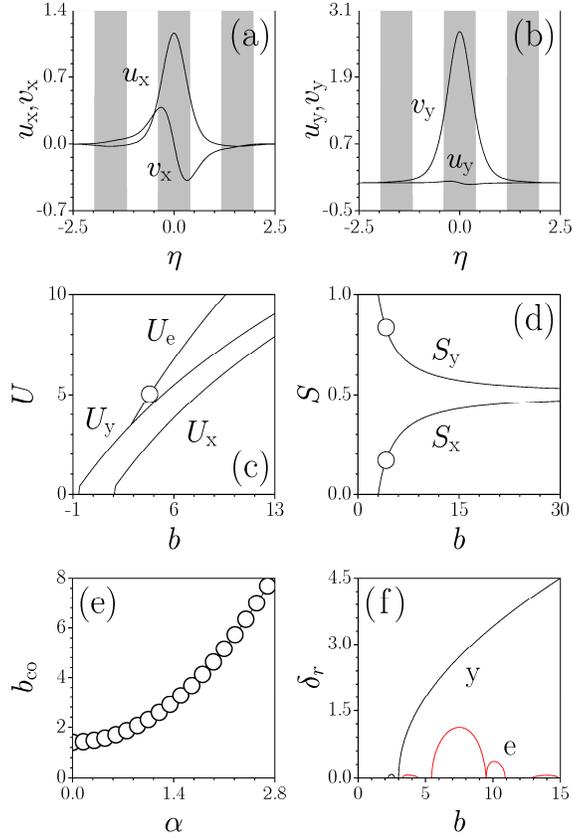

Figure 1. Profiles of (a) $x$ and (b) $y$ components of elliptically polarized vector soliton at $b = 4.2$, $\alpha = 1.4$. In gray regions $R(\eta) \geq 0$; in white regions $R(\eta) < 0$. (c) $U$ versus $b$ for $x$-, $y$-, and elliptically polarized solitons at $\alpha = 1.4$. (d) $S_x$ and $S_y$ versus $b$ at $\alpha = 1.4$. Points in (c), (d) correspond to solitons in (a), (b). (e) Cutoff for existence of elliptically polarized soliton versus $\alpha$. (f) $\delta_r$ versus $b$ for $y$- and elliptically polarized solitons at $\alpha = 1.4$. In all cases $p = 4$.

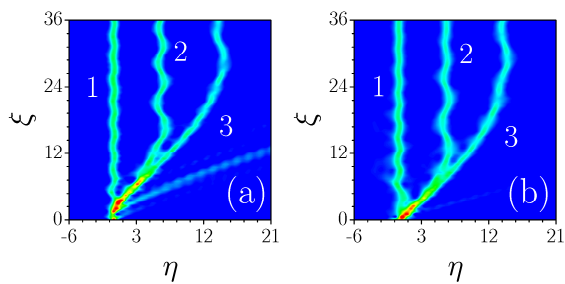

Figure 2. (a) Dynamics of $x$ component at $p = 0.50$ (1), $0.36$ (2), and $0.22$ (3) when input beam corresponds to vector soliton obtained at $b = 6$, $p = 4$. (b) The same as in (a) but for sech-shaped input beams and $p = 0.60$ (1), $0.32$ (2), and $0.23$ (3). Distributions corresponding to different $p$ are superimposed. In all cases $\alpha = 1.4$.

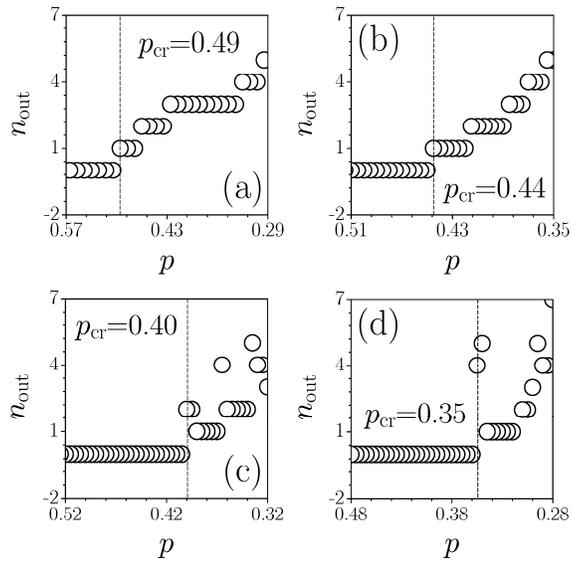

Figure 3. The number of the output channel versus $p$ at $\alpha = 1.4$ for elliptically polarized input vector soliton with (a) $U = 6.38$, (b) $6.78$, (c) $6.98$, and (d) $7.15$ obtained at $p = 4$. Vertical dashed lines indicate critical lattice depth.

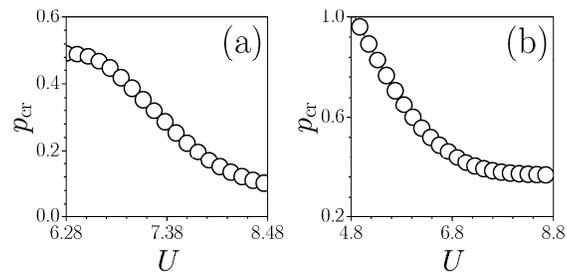

Figure 4.     $p_{\rm cr}$ versus $U$ for (a) elliptically polarized input vector soliton and (b) sech-shaped input beams at $\alpha = 1.4$.